\documentclass[12pt]{iopart}
\begin{document}

\title{Alternating current driven instability in magnetic junctions}

\author{E M Epshtein and P E Zilberman}

\address{V A Kotelnikov Institute of Radio Engineering and Electronics
of the Russian Academy of Sciences, Fryazino, 141190, Russia}
\ead{epshtein36@mail.ru}
\begin{abstract}
An effect is considered of alternating (high-frequency) current on the
spin-valve type magnetic junction configuration. The stability with
respect to small fluctuations is investigated in the macrospin approximation.
When the current frequency is close to the eigenfrequency (precession frequency) of the free layer,
parametric resonance occurs. Both collinear configurations, antiparallel
and parallel ones, can become unstable under resonance conditions.
The antiparallel configuration can become unstable under non-resonant conditions,
also. The threshold current density amplitude is of the order of the dc current
density switching the magnetic junction.
\end{abstract}



\submitto{\JPCM}

%\maketitle

\section{Introduction}\label{section1}
There is unremitting attention to the behaviour of magnetic junctions
under spin-polarized current flowing. It is not surprising because of
a number of interesting phenomena which have been observed, such as magnetic
configuration switching~\cite{Katine}, spin wave generation~\cite{Tsoi},
etc. The effects can occur on a nanosize scale, since their
characteristic lengths are the exchange and spin diffusion ones with
typical values of the order of 10 nm. This allows to use such effects for
high-density information recording by electric current, unattainable for
switching magnetic elements by magnetic field alone.

The current-driven switching of magnetic junctions is accompanied often with
magnetization oscillations and the other high-frequency effects (see,
e.g.,~\cite{Tsoi}--\cite{Xiao}). In this connection, an interesting
problem arises, namely, effect of spin-polarized {\it alternating} current
on magnetic junctions.

In this work, we consider an effect of alternating (high-frequency)
current on the magnetic junction configuration. When the parametric
resonance conditions fulfill, both collinear configurations, parallel and
antiparallel ones, can become unstable. It should be noted that the
parametric resonance in magnetics was studied in many works (see,
e.g.,~\cite{Gurevich}). However, the parametric resonance was considered
there which was excited by high-frequency external magnetic field, i.e. the nonlocal Ampere
field. In our case, the spin-polarized current interaction with magnetic
lattice is of exchange nature, so that it is localized in the
above-mentioned small range. As will be shown below, the current-induced
parametric resonance has additional features. Thus, the main effect takes
place at the precession frequency, not the doubled one. The instability is possible, too,
beyond the parametric resonance conditions. It appears that only the
spin-injection mechanism~\cite{Heide,Gulyaev1}, not the spin-torque
transfer (STT) mechanism~\cite{Slonczewski,Berger}, contributes to the effects in consideration.

\section{Model considered and main equations}\label{section2}
We consider a conventional spin-valve model consisted of a pinned
ferromagnetic layer (layer 1), thin spacer layer, ferromagnetic free layer (layer 2) and
nonmagnetic layer (layer 3) closing electric circuit. The alternating current flows perpendicular to the
layer planes (CPP mode). We investigate stability of collinear (parallel or antiparallel)
relative orientation of the pinned and free layers against small
magnetization fluctuations under alternating current flowing with density
\begin{equation}\label{1}
  j(t)=j_0\cos\Omega t.
\end{equation}
The free layer is
assumed to be thin compared to spin diffusion length and domain wall
thickness, so that the macrospin approximation is
applicable~\cite{Gulyaev2}. In this approximation, the fluctuations are
described by the modified Landau--Lifshitz--Gilbert (LLG) equation~\cite{Gulyaev2}
\begin{eqnarray}\label{2}
  \frac{\rmd\hat{\bi M}}{\rmd t}-\kappa\left[\hat{\bi M}\times\frac{\rmd\hat{\bi
  M}}{\rmd t}\right]+\gamma\left[\hat{\bi M}\times\bi H\right]\nonumber\\
  +\gamma H_{\rm{a}}\left(\hat{\bi M}\cdot\bi{n}\right)\left[\hat{\bi M}\times\bi{n}\right]
  +\gamma\left[\hat{\bi M}\times\bi{H}_{\rm{d}}\right]\nonumber\\
  +\frac{ap}{L}\left[\hat{\bi M}\times\hat{\bi M}_1\right]
  +\frac{ak}{L}\left[\hat{\bi M}\times\left[\hat{\bi M}\times\hat{\bi
  M}_1\right]\right]=0.
\end{eqnarray}
Here the following notations are used: $\hat\bi M=\bi M/|\bi M|$ is unit vector
along the free layer magnetization $\bi M$, $\hat{\bi M}_1$ is the same for the
pinned layer, $\bi{H}$ is external magnetic field,
$H_{\rm{a}}$ is anisotropy field, $\bi n$ is unit vector along the
anisotropy axis, $\bi{H}_{\rm{d}}$ is demagnetization field, $L$ is the
free layer thickness, $a$ is the magnetization diffusion constant,
$\gamma$ is gyromagnetic ratio, $\kappa$ is the Gilbert damping constant.
The parameters~$p$ and~$k$ describe the spin-polarized current effect on
the free layer magnetic lattice due to spin-injection
mechanism~\cite{Heide,Gulyaev1} and STT mechanism~\cite{Slonczewski,Berger},
respectively. In collinear configuration
with~$\left(\hat{\bi M}_1\cdot\hat{\bi M}\right)=\pm1$, these
parameters take the following forms:
\begin{equation}\label{3}
  p=\frac{\mu_{\rm{B}}\gamma\alpha\tau Q_1}{ea}\frac{\lambda\nu(\nu^\ast-1)
  \pm2b\nu^\ast}{(\nu^\ast+1)^2}|j|\equiv p^{(\pm)},
\end{equation}

\begin{equation}\label{4}
  k=\frac{\mu_{\rm{B}}Q_1}{eaM}\frac{\nu^\ast}{\nu^\ast+1}j,
\end{equation}
where $e$ is electron charge, $\mu_{\rm B}$ is the Bohr magneton, $Q_1$ is the conductivity spin
polarization in layer 1, $\lambda=L/l\ll1$, $l$ being the spin
diffusion length in the free layer, $\tau$ is spin relaxation time in the free layer,
$\alpha$ is the dimensionless {\it sd} exchange interaction constant in the free layer,
$b=(\alpha_1M_1\tau_1)/(\alpha M\tau)$ ratio describes the pinned layer contribution
(see~\cite{Epshtein1} for more details), $\nu=Z_1/Z_2$, $\nu^\ast=\lambda\nu+(Z_1/Z_3)$;
$Z_i\,(i=1,\,2,\,3)$ being the spin resistances~\cite{Epshtein2}
\begin{equation}\label{5}
  Z_i=\frac{l_i}{\sigma_i\left(1-Q_i^2\right)},
\end{equation}
$\sigma_i$ is conductivity of the $i$-th layer. The upper and lower signs
in~\eref{3} refer to the parallel and antiparallel
configurations, respectively.

The formulae~\eref{2}--\eref{4} have been derived on the assumption of
direct current flowing. However, when the alternating current frequency
$\Omega$ is comparable with the magnetization precession frequency, the
conduction electrons can follow the magnetization oscillations, so that
all the conditions fulfill under which the formulae are valid. Therefore,
we may substitute the alternating current density~\eref{1} with frequency
$\Omega$ for $j$ in formulae~\eref{3} and~\eref{4}. Then the parameters
$p$ and $k$ in~\eref{2} will have time dependence of the form
$p^{(\pm)}(t)=p_0^{(\pm)}|\cos\Omega t|$ and $k(t)=k_0\cos\Omega t$, respectively.

Note that the contribution of the spin-injection mechanism is proportional
to the absolute value of the current density, so that it is the same for
forward ($1\to2\to3$) and backward ($3\to2\to1$) currents, in contrast
with the contribution of the STT mechanism. This leads to
different spectra of $p^{(\pm)}(t)$ and $k(t)$ functions:
\begin{eqnarray}\label{6}
  p^{(\pm)}(t)=p_0^{(\pm)}|\cos\Omega t|\nonumber\\
  =p_0^{(\pm)}\left(\frac{2}{\pi}+\frac{4}{3\pi}\cos2\Omega
  t-\frac{4}{15\pi}\cos4\Omega t+\ldots\right),
\end{eqnarray}
i.e., only the even harmonics of $\Omega$ frequency (including dc component) are presented in the
$p^{(\pm)}(t)$ spectrum, while the $k(t)$ function spectrum consists of the single
frequency $\Omega$. Therefore, two mechanisms do not interplay in the
lowest-order resonance phenomena and they may be considered separately. We
begin with the spin-injection mechanism.

Let $x$ axis be directed along the current, $yz$ plane be parallel to the
layer planes, the free layer occupy the range $0\le x\le L$, vectors $\bi
H,\,\bi n$ and $\hat{\bi M}_1$ have the following components: $\bi
H=\{0,\,0,\,H_z\}$, $\bi n=\{0,\,0,\,1\}$, $\hat{\bi M}_1=\{0,\,0,\,1\}$.
We investigate stability of equilibrium collinear configurations
$\bar{\hat{M}}_z=\pm1$ against the free layer magnetization small
fluctuations $\hat M_x,\,\hat M_y$. The LLG equation components linearized in the
fluctuations taking the spin-injection mechanism into account only, take
the form
\begin{eqnarray}\label{7}
  \frac{\rmd\hat M_x}{\rmd t}+\kappa\bar{\hat M}_z\frac{\rmd\hat M_y}{\rmd t}+
  \gamma(H_z+H_{\rm{a}}\bar{\hat M}_z)\hat M_y\nonumber\\
  +\frac{ap^{(\pm)}(t)}{L}\hat M_y=0,
\end{eqnarray}

\begin{eqnarray}\label{8}
  \frac{\rmd\hat M_y}{\rmd t}-\kappa\bar{\hat M}_z\frac{\rmd\hat M_x}{\rmd t}-
  \gamma(H_z+H_{\rm{a}}\bar{\hat M}_z+4\pi M\bar{\hat M}_z)\hat M_x\nonumber\\
  -\frac{ap^{(\pm)}(t)}{L}\hat M_x=0.
\end{eqnarray}
The periodic time dependence of the coefficients of the last terms in the left-hand
side of the equations leads to possibility of parametric resonance.

\section{Parametric resonance}\label{section3}
It is well known~\cite{Landau}, that the parametric resonance occurs when
the parameter modulation frequency is close to $2\omega_0/n$,
where $\omega_0$ is eigenfrequency of the oscillatory system,
$n=1,\,2,\,3,\,\ldots$. If the modulation coefficient is small, the
parametric instability range narrows and the instability threshold rises
with increasing the resonance order $n$.

In accordance with~\eref{6}, we consider parametric excitation at
frequency $2\Omega$ with the first two terms taking into account in the
right-hand side of~\eref{6}. We assume the damping constant $\kappa$ to be
small and neglect it for the time. Taking the Fourier transforms
of~\eref{7},~\eref{8}  with respect to time, we have
\begin{eqnarray}\label{9}
    -\rmi\omega\hat{M}_x(\omega)+\gamma\left(H_z+H_{\rm
    a}+\frac{3\epsilon}{\gamma}\right)\hat{M}_y(\omega)\nonumber\\
    =-\epsilon^{(\pm)}\left[\hat{M}_y(\omega+2\Omega)+\hat{M}_y(\omega-2\Omega)\right],
\end{eqnarray}
\begin{eqnarray}\label{10}
    -\rmi\omega\hat{M}_y(\omega)-\gamma\left(H_z+H_{\rm
    a}+4\pi M\bar{\hat{M}}_z+\frac{3\epsilon}{\gamma}\right)\hat{M}_x(\omega)\nonumber\\
    =\epsilon^{(\pm)}\left[\hat{M}_x(\omega+2\Omega)+\hat{M}_x(\omega-2\Omega)\right];
\end{eqnarray}
\begin{equation}\label{11}
  \epsilon^{(\pm)}=\frac{2}{3\pi}\frac{ap_0^{(\pm)}}{L}
\end{equation}
is a quantity proportional to the current density amplitude with
dimension of frequency.

Usually, a condition is fulfilled
\begin{equation}\label{12}
  4\pi M\gg|H_z|,\,H_{\rm a},\,3\epsilon^{(\pm)}/\gamma,
\end{equation}
that is assumed below.

If we make the substitution $\omega\to\omega\pm2\Omega$
in~\eref{10},~\eref{11}, the equations are obtained where terms with
$\hat{M}_{x,\,y}(\omega\pm2\Omega)$ stand in the left-hand sides and the
terms with $\hat{M}_{x,\,y}(\omega)$ and
$\hat{M}_{x,\,y}(\omega\pm4\Omega)$ with $\epsilon^{(\pm)}$ coefficient in the
right-hand sides.

If frequency $\epsilon^{(\pm)}$ (the coupling parameter) is small compared to the other characteristic
frequencies ($\omega_0,\,\Omega$), the coupling with
$\hat{M}_{x,\,y}(\omega\pm4\Omega)$ can be neglected, because it leads to
higher-order corrections in $\epsilon^{(\pm)}$. As a result, we obtain a closed
system of equations for $\hat{M}_{x,\,y}(\omega)$ and
$\hat{M}_{x,\,y}(\omega\pm2\Omega)$. Equating the determinant of the
system to zero, we get the dispersion equation
\begin{eqnarray}\label{13}
  \left\{\omega+(\epsilon^{(\pm)})^2\left[\frac{\omega+2\Omega}{\Delta(\omega+2\Omega)}+
\frac{\omega-2\Omega}{\Delta(\omega-2\Omega)}\right]\right\}^2-\omega_0^2\nonumber\\
+(\epsilon^{(\pm)})^2(4\pi\gamma M)^2\left[\frac{1}{\Delta(\omega+2\Omega)}+
\frac{1}{\Delta(\omega-2\Omega)}\right]\nonumber\\
-(\epsilon^{(\pm)})^4\omega_0^2\left[\frac{1}{\Delta(\omega+2\Omega)}+
\frac{1}{\Delta(\omega-2\Omega)}\right]^2=0,
\end{eqnarray}
where
\begin{equation}\label{14}
  \omega_0=2\gamma\left[\pi M\left(H_z\bar{\hat{M}}_z+H_{\rm
  a}+\frac{3\epsilon^{(\pm)}}{\gamma}\bar{\hat{M}}_z\right)\right]^{1/2}
  \end{equation}
  is the eigenfrequency neglecting the terms quadratic in $\epsilon^{(\pm)}$,
\begin{equation}\label{15}
  \Delta(\omega)=\omega_0^2-\omega^2.
\end{equation}
Note, that the dc component of $p^{(\pm)}(t)$ (see~\eref{6}) renormalizes the
system eigenfrequency.

Near the parametric resonance, we have
$\omega\approx\omega_0\approx\Omega$, so that we may retain only the
summands with resonant denominator $\Delta(\omega-2\Omega)$ in the terms
with $\epsilon^{(\pm)}$ and replace $\omega$ and $\Omega$ with $\omega_0$
everywhere but that denominator. This leads to the parametric resonance
equation
\begin{equation}\label{16}
  \Delta(\omega)\Delta(\omega-2\Omega)=(4\pi\gamma M\epsilon^{(\pm)})^2.
\end{equation}
Let $\omega=\omega_0+\nu$, $\Omega=\omega_0+\delta$, where
$|\nu|,\,|\delta|\ll\omega_0$. The equation for $\nu$ takes the form
\begin{equation}\label{17}
  \nu^2-2\delta\nu+\left(\frac{2\pi\gamma M\epsilon^{(\pm)}}{\omega_0}\right)^2=0,
\end{equation}
which gives
\begin{equation}\label{18}
  \nu=\delta\pm\left[\delta^2-\left(\frac{2\pi\gamma
  M}{\omega_0}\right)^2(\epsilon^{(\pm)})^2\right]^{1/2}.
\end{equation}
It is seen from~\eref{18} that the parametric instability takes place at
\begin{equation}\label{19}
  \epsilon^{(\pm)}>\frac{\omega_0\delta}{2\pi\gamma M}
\end{equation}
with increment
\begin{equation}\label{20}
  (\mathrm{Im}\,\omega)_{\rm p}=\left[\left(\frac{2\pi\gamma
  M\epsilon^{(\pm)}}{\omega_0}\right)^2-\delta^2\right]^{1/2}.
\end{equation}
In presence of dissipation ($\kappa\ne0$), damping takes place with
decrement (see, e.g.,~\cite{Gurevich})
\begin{equation}\label{21}
  |(\mathrm{Im}\,\omega)_{\rm d}|=2\pi\kappa\gamma M.
\end{equation}
If $\kappa\ll1$, the parametric instability threshold may be estimated from
the condition
\begin{equation}\label{22}
    (\mathrm{Im}\,\omega)_{\rm p}>|(\mathrm{Im}\,\omega)_{\rm d}|
\end{equation}
at zero resonance detuning ($\delta=0$).

Equations~\eref{20}--\eref{22} give the following condition for the parametric
instability:
\begin{equation}\label{23}
  \epsilon^{(\pm)}>\kappa\omega_0.
\end{equation}
The right-hand side of this inequality contains $\epsilon^{(\pm)}$, too.
Therefore, the inequality is to be resolved with respect to $\epsilon^{(\pm)}$. As
a result, the following instability threshold is obtained for the
collinear configurations
\begin{equation}\label{24}
  \epsilon_{\rm th}^{(\pm)}=6\pi\kappa^2\gamma M\left[\left(\frac{H_{\rm a}\pm
  H_z}{9\pi\kappa^2M}+1\right)^{1/2}\pm1\right].
\end{equation}

In contrast with the case of direct current, both collinear configurations
can become unstable, but the corresponding thresholds are different. At
$\epsilon^{(-)}>\epsilon_{\rm th}^{(-)}$, $\epsilon^{(+)}<\epsilon_{\rm
th}^{(+)}$ the switching from unstable antiparallel configuration to stable parallel
one is possible, while at $\epsilon^{(+)}>\epsilon_{\rm th}^{(+)}$ both
configurations are unstable. At $H_{\rm a}\pm H_z\gg9\pi\kappa^2M$ the
instability threshold for both configurations takes the form
\begin{equation}\label{25}
    \epsilon_{\rm th}^{(\pm)}=2\kappa\gamma[\pi M(H_{\rm
    a}\pm H_z)]^{1/2}=\kappa\omega_0^{(0)},
\end{equation}
where $\omega_0^{(0)}=2\gamma[\pi M(H_{\rm a}\pm H_z)]^{1/2}$ is the eigenfrequency
in absence of the electric current.

If $H_{\rm a}-H_z\ll9\pi\kappa^2M$, we have
\begin{equation}\label{26}
    \epsilon_{\rm th}^{(-)}=\frac{1}{3}\gamma(H_{\rm a}-H_z)
\end{equation}
for the antiparallel configuration. It is seen from~\eref{26}, that the
instability threshold can be lowered considerably with external magnetic field
close to, but slightly lower than the anisotropy field. Note, that
participation of the magnetic field does not prevent locality of the
effect, because the magnetic field cannot do switching alone, without the
current.

Let us compare the alternating current density amplitude $j_{\rm th}$ corresponding to the
parametric instability threshold with the the direct current density $j_{\rm dc}$ leading to the
switching antiparallel orientation to parallel one in absence of the
external magnetic field. The dc threshold corresponds
to the condition~\cite{Epshtein1} $ap^{(-)}/L>\gamma H_{\rm a}$, where
$p^{(-)}$ is determined by~\eref{3}. In the parametric resonance case,
$ap_{\rm th}^{(-)}/L=3\pi\kappa\gamma(\pi MH_{\rm a})^{1/2}$, so that
\begin{equation}\label{27}
  \frac{j_{\rm th}}{j_{\rm dc}}=3\pi\sqrt\pi\kappa\left(\frac{M}{H_{\rm
  a}}\right)^{1/2}.
\end{equation}
At typical values of the parameters ($M/H_{\rm
a}\sim10,\,\kappa=3\times10^{-2}$), this ratio is of the order of 1. At
lower damping constant, the parametric instability threshold will be
smaller than the dc threshold.

\section{Non-resonance instability}\label{section4}
The instability of the antiparallel configuration under alternating current
flowing is possible also when the parametric resonance condition
$\Omega\approx\omega_0$ does not fulfill. It follows from~\eref{14} that
the eigenfrequency $\omega_0$ becomes imaginary at $\epsilon^{(-)}>\gamma(H_{\rm
a}-H_z)/3$ because of the contribution of dc component in the spectrum of $p^{(\pm)}(t)$
function, i.e., such a component of the {\it sd} exchange field. The
cancelling of the eigenfrequency corresponds to an orientational phase
transition similar to that under dc injection current~\cite{Epshtein1}.
The threshold amplitude of the alternating current is $\pi/2$ times as
much as the corresponding dc threshold. Note, that the threshold is higher
then the parametric instability threshold $\epsilon_{\rm th}^{(-)}$, so
that the parametric instability develops first under fulfilled parametric
resonance conditions.

\section{Is similar effect possible due to the spin-torque transfer mechanism?}\label{section5}
Since the $k(t)$ function describing the STT mechanism contribution has
a single-mode spectrum, the lowest order of the parametric resonance
corresponds to $\Omega\approx2\omega_0$ condition. However, if all the
previous calculations are carried out for $p^{(\pm)}(t)=0,\,k(t)\ne0$, the
contributions from $k(t)$ mutually cancel in resonance
approximation ($\Omega\approx2\omega\approx2\omega_0$). Therefore, a
negative answer should be given to the question stated above. This is due to the
fact, that the STT, in contrast with the spin injection, modifies the
damping, not the eigenfrequency. As is known, the parametric resonance is
not possible in such a situation.

Such a difference between two mechanisms can be used for separating their
contributions in various cases.

\section{Conclusion}\label{section7}
The analysis carried out shows that the collinear configurations of a
magnetic junction can become unstable under parametric resonance
conditions. The instability threshold is of the same order of magnitude as
the corresponding threshold under direct current flowing through the
magnetic junction. Depending on the alternating current density amplitude,
the antiparallel configuration can be switched to parallel one or both
collinear configurations can be unstable with growing magnetic
fluctuations. To elucidate the resulting state which the system considered comes
to, further investigations, both theoretical and experimental, are needed.

\ack
The authors are grateful to Dr.~S G Chigarev and Dr.~A I Krikunov
for helpful discussion.

The work was supported by the Russian Foundation for Basic Research, Grants Nos. 06-02-16197
and 08-07-00290.

\end{document}